\documentclass[10pt,a4paper]{article}

\title{On seat allocation problem with multiple merit lists}

\author{Rahul Kumar Singh {\normalsize{and }} Sanjeev
Saxena\thanks{E-mail: ssax@cse.iitk.ac.in}\\ Dept. of Computer Science
and Engineering,\\ Indian Institute of Technology,\\ Kanpur, INDIA-208
016}

\date{} 

\begin{document}
\maketitle

\subsection*{\centering{Abstract}}

In this note, we present a simpler algorithm for joint seat allocation
problem in case there are two or more merit lists. In case of two
lists (the current situation for Engineering seats in India), the
running time of the algorithm is proportional to sum of running time
for two separate (delinked) allocations. The algorithm is straight
forward and natural and is not (at least directly) based on deferred
acceptance algorithm of Gale and Shapley. Each person can only move
higher in his or her preference list. Thus, all steps of the algorithm
can be made public. This will improve transparency and trust in the
system.

\section{Introduction}

In India seats in Engineering colleges are filled based on two
different tests--- JEE (advanced) and JEE (main). JEE (advanced) merit
lists is used for courses in I.I.T.s (Indian Institute of Technology)
and JEE (main) list is used for courses in other centrally funded
colleges. First preference of most students is some popular course
(currently Computer Science) in an I.I.T.. 
After the two merit lists are prepared, students get a window of few
days to fill up their individual combined preferences. Thus, the time
taken to input individual preference lists by students can be ignored.

There is also reservation based on social and economic criteria. A
fraction of seats in each course is reserved, and these seats (for all
practical purposes) can only be filled by persons of that category.
However, in case persons in reserved category do better, then they can
also opt for general or unreserved seats. 

In addition, to ensure that number of female candidates are adequately
represented, additional supernumerary seats are created. Thus, if
number of female students who get admission in a course is $x$ instead
of desired number $y$, $\min\{0,y-x\}$ extra seats are created. These
can only be filled by female candidates.

Baswana et.al.[1,2] have described a method to implement this
based on the ``deferred acceptance'' (DA) algorithm of Gale and
Shapley[3], with some ad hoc heuristics to take care of
supernumerary seats. 

In this note, we present a simpler algorithm for joint seat allocation
problem in case there is more than one merit list. In case of two
lists, 
the running time of the algorithm is proportional to sum of running
time for two separate (delinked) allocations. The algorithm is
straight forward and natural and is not (at least directly) based on
differed allocation algorithm of Gale and Shapley[3]. In
particular, there is no ``deallocation'' (only empty seats or seats
which become empty are filled). Each person can only move higher in
his or her preference list (i.e., it is monotone). Thus, all steps of
the algorithm can be made public. This will improve transparency and
trust in the system.

The problem of more than two merit lists is discussed in Section~3. We
look at issues associated with reservation in Section~3.1. 
Supernumerary seats are discussed in Section~3.2. We ignore the
problem of ties, for simplicity.

\subsection{Simpler Problems}

Let us first look at a simpler problem, when there is only one merit
list, say $L$. The ``obvious'' algorithm is:\\
~\\
Look at each person in turn from the best ranked person to the worst
in $L$
\begin{quote}
If $i$th person is being looked at, we look at his/her preferences
(from most preferred course to the worst) and assign the first course
which has not been completely filled.
\end{quote}
If $i$th person gets $p_i$th preference, then the time taken by the
algorithm is $O(m+\sum p_i)$; here $m$ is the total number of courses.
We are assuming that there may be courses which no person is
interested in.

Let us, next consider the case, when there are two merit lists, say
$L1$ and $L2$, but students are allocated courses independently. If
$i$th person gets preference $p_i$ based on $L1$ and gets preference
$q_i$ based on list $L2$. The time taken for the first allocation
(using an algorithm similar to the one list case) will be $O(m_1+\sum
p_i)$; here $m_1$ is the total number of courses in which admission is
based on $L1$. And the time for the second allocation will be
$O(m_2+\sum q_i)$; here $m_2$ is the total number of courses in which
admission is based on $L2$. Or, the total time is $T=O(m+\sum p_i+\sum
q_i)$, where $m=m_1+m_2$.

\section{Joint Allocation}

Next let us look at the original problem. Without loss of generality,
we assume that most candidates have highest preferences for colleges
in $L1$ list (say popular courses in I.I.T.s where admission is made
based on JEE (advanced)). 

At high level the algorithm is in two steps:
\begin{itemize}
\item Do allocation based on the first list (ignoring the
preferences based on the other list).
\item Then look at persons, in order of merit, in the second list. If any
person can get a more preferred course from the second list, then that
person is assigned that course. His seat is offered to the next (in
order of merit) interested person; the process is repeated for the
newly created vacant
seat. 
\end{itemize}
Let us look at each step in more detail.

\subsection{First Step}

The first step is basically doing allocation using only list $L1$ (for
colleges which use list $L1$), but with some additional book-keeping. \\
~\\
for each $L1$-list rank $i$  in turn (from best to worst do)\\

We go down the $i$th list only looking at courses where admission is
done on basis of $L1$.
\begin{enumerate}
\item If a course at $j$th preference is completely filled, we put
$i$ in the waiting list of $j$th course. Each waiting list is a queue
(first in first out, usual queue). 

\item Else, we allot the $j$th course (say one at $p_i$th position) and
look at the next person.
\end{enumerate}

The time taken for $i$th person is still $O(p_i)$. 

Remark~1. If $i$th person gets $p_i$th preference, he is added in
waiting lists of courses which are his $1,2,{\ldots} ,p_i-1$ preference.

Remark~2. Instead of storing full record for each person in the queue
(waiting lists), we only store a pointer to that person. 

Remark~3. For each
person, we also store the preference number which is currently allotted.
Thus, we can find the course allotted to a person in $O(1)$ time.

Remark~4. When a person is inputting his/her preference, in addition
to storing the combined preference, we also store the $L1$ and $L2$
preference in separate lists (in addition to combined priority lists).

Eg: 
$
\begin{array}{|l|l|l|l|l|l|l|l|}
\hline
\mbox{Preference} & 1 & 2 & 3 &4 & 5 & 6 & 7\\
\hline
Course & A & B & C &D & E & F & G\\
\hline
L1 \mbox{ or } L2 & 1 & 1 & 1 &2 & 1 & 2 & 2 \\
\hline
\end{array}
$

Then we store the first two lines in joint preference list and also
store (preference in list, overall preference number, course):

L1 list: (1,1,A), (2,2,B), (3,3,C), (4,5,E)

L2 list: (1,4,D), (2,6,F), (3,7,G)

Hence, in Step~1, no time is ``wasted'' in looking at preferences in $L2$.

\subsection{Second Step}

In the second step we check if some $L2$ list course has higher
preference than the course currently allotted.\\
~\\
for each $L2$-list rank $i$  in turn (from best to worst do)\\

We go down the $i$th list
\begin{enumerate}
\item If first $L2$ list preference is lower (less preferred) than the
course allotted, look at the next person in the list.

\item If a course at $j$th preference is completely filled, we put
$i$ in waiting list for the $j$th course. Each waiting list is again a
queue. 

Remark: This step is required in case there are more than two lists,
or when we allow a student to withdraw, otherwise, this step is not
required.

\item Else, if the $j$th preference course is more preferred than
the course currently allotted we allot the $j$th preference
course(say $q_i$). 

Remark: Time for this step is again $O(q_i)$.

\item The earlier $L1$-list course of $i$ is offered  to the first person
(say $c_1$) in the waiting list for that course (say $D_1$). In case,
the course currently allotted to $c_1$ is more preferred (by $c_1$),
we let $c_1$ be the next person in the waiting list of $D_1$. The
process is repeated until either we find a person (lets us also call
him $c_1$) in waiting list of $D_1$ who wants to take $D_1$, or the
waiting list gets exhausted.

Course earlier allotted to $c_1$ (say $D_2$) is similarly offered to
the first person (say $c_2$) in waiting list for $D2$, and so on,
until a person who did not had a course allotted is encountered, or
there is no person in waiting list for that course.

Remark: As we are only going down on each waiting list, the time taken
over the entire algorithm can not be more than the sum of length of
all waiting lists. If a person $i$ gets course of priority $p_i$ in
Step~1, $i$ is in $(p_i-1)$ waiting lists. Or total length of all
waiting lists is $O(\sum p_i)$.
\end{enumerate}

As each person after reallocation gets a course which is higher in
his/her preference list, there are no cycles. 
Time taken for Step 2 is $O(m+\sum q_i+\sum p_i)$.
Or total time is $O(T)$, the same (up to a constant multiplicative
factor) as for two separate allocations.

For correctness, observe that as allocation in first step is done in
order of merit, a person with lower rank cannot get what a higher ranked
person failed to get.

In second step, as waiting lists are in decreasing order of merit,  a
vacant seat will be first assigned to the highest ranked person (who
could not get it).

Remark: The method can also be used in case, we 
permit a student to withdraw. If a person withdraws, the next person
in the waiting list is offered that course.

\section{More than two lists}

Let us next consider the case when there are more than two lists.
Assume that there are three lists. We run the algorithm of Section~2
based on first two lists and do the allocation. Then in third step
(which will be same as the second step), we use the third list (say)
$L3$ instead of $L2$.

For correctness, we have already seen that after allocation in second
step, a person with lower rank in list $L1$ (respectively, $L2$)
cannot get what a course which a person higher ranked in $L1$ ($L2$)
failed to get.

In third step, as waiting lists are in decreasing order of merit, a
vacant seat will be first assigned to the highest ranked person in
appropriate merit list (who could not get it).

The process can clearly be generalised to more than three lists.

\subsection{Reservation}

We will assume that set of persons in any reserved list form an
ordered subsequence of persons in the full (un-reserved) list. Thus,
if person $A$ is better ranked than $B$ in $L1$-general or unreserved
list, then $A$ is also better ranked than $B$ in the $L1$-reserved list.

We replace each preference of persons in reserved list by a pair of preferences:

(preference $i$, course $j$) is replaced by pairs
\begin{quote}
(preference $2i-1$, course $j$, unreserved) and\\
(preference $2i$, course $j$, reserved)
\end{quote}

We then run the algorithm as before. It is straight forward to take
care of nested reservations (like physically challenged students in
reserved category).

\subsection{Supernumerary Seats}

In step 1, when a course gets filled, we add necessary additional
supernumerary seats (say based on gender). As in each iteration, we only
need to know whether a seat is still vacant or not (and not the total
number of vacant seats), supernumerary seats can be treated as
``reservation'' and number of seats in the category can change as the
algorithm progresses.

There are some obvious changes. If in Step~2, a female candidate in
gender-neutral category vacates a seat, an additional supernumerary
seat has to be created. This seat is also filled as before.

We can also use the heuristics proposed in [1,2]

\subsection*{Disclaimer}

Neither author is associated with seat allocation process in India.

\section*{References}

\begin{enumerate}
\item  S.Baswana, P.P.Chakrabarti, S.Chandran, Y.Kanoria and U.
Patange, Centralized Admissions for Engineering Colleges in India.
Interfaces 49(5): 338-354 (2019)

\item  S.Baswana, P. P. Chakrabarti, Y.Kanoria, U.Patange, S.
Chandran, Joint Seat Allocation 2018: An algorithmic perspective. CoRR
abs/1904.06698 (2019)

\item  D.Gale and L.S.Shapley, College admissions and the
stability of marriage, Amer. Math. Monthly, 69(1): 9-15 (1962).

\end{enumerate}
\end{document}